\documentclass[conference]{IEEEtran}
\IEEEoverridecommandlockouts

\usepackage{cite}
\usepackage{amsmath,amssymb,amsfonts}
\usepackage{algorithmic}
\usepackage{graphicx}
\usepackage{textcomp}
\usepackage{xcolor}
\usepackage{tabularx} 
\usepackage{eso-pic}  

\newcommand{\PlaceAcceptedManuscriptNotice}{%
  \AddToShipoutPictureFG*{%
    \AtPageLowerLeft{%
      \raisebox{\dimexpr 0.08in+10mm\relax}[0pt][0pt]{%
        \hspace*{\dimexpr 0.08in+15mm\relax}%
        \parbox[b]{\dimexpr\paperwidth-0.16in-15mm\relax}{%
          \raggedright
          \fontsize{6.5pt}{6.5pt}\selectfont
          \textbf{\textit{Accepted manuscript.}}\\[-0.2pt]
          \textcopyright~2025 IEEE. Personal use of this material is permitted. Permission from IEEE must be obtained for all other uses, in any current or future media, including reprinting/republishing this material for\\[-0.2pt]
          advertising or promotional purposes, creating new collective works, for resale or redistribution to servers or lists, or reuse of any copyrighted component of this work in other works.\\[-0.2pt]
          \textit{Version of Record:} K. Tamesue et al., ``Experimental Verification on LoS-MIMO Transmission in 300-GHz Band,'' in \textit{2025 IEEE Conference on Antenna Measurements and Applications (CAMA)},\\[-0.2pt]
          Antibes Juan-les-Pins, France, 2025, doi: 10.1109/CAMA65664.2025.11335203.%
        }%
      }%
    }%
  }%
}


\def\BibTeX{{\rm B\kern-.05em{\sc i\kern-.025em b}\kern-.08em
    T\kern-.1667em\lower.7ex\hbox{E}\kern-.125emX}}
\begin{document}

\title{Experimental Verification on LoS-MIMO Transmission in 300-GHz Band\\
}

\author{
\centering
\IEEEauthorblockN{Kazuhiko Tamesue}
\IEEEauthorblockA{\textit{Faculty of Science and Engineering} \\
\textit{Waseda University}\\
Tokyo, Japan \\
ktamesue@aoni.waseda.jp}
\and
\IEEEauthorblockN{Seiji Nishi}
\IEEEauthorblockA{\textit{Faculty of Science and Engineering} \\
\textit{Waseda University}\\
Tokyo, Japan \\
s.nishi3@kurenai.waseda.jp}
\and
\IEEEauthorblockN{Kunihisa Jitsuno}
\IEEEauthorblockA{\textit{Faculty of Science and Engineering} \\
\textit{Waseda University}\\
Tokyo, Japan \\
k.jitsuno@kurenai.waseda.jp}
\and
\IEEEauthorblockN{\hspace{0.1cm}Toshio Sato}
\IEEEauthorblockA{\hspace{0.8cm}\textit{Faculty of Science and Engineering} \\
\textit{Waseda University}\\
Tokyo, Japan \\
toshio4.sato@aoni.waseda.jp}
\and
\IEEEauthorblockN{\hspace{0.5cm}Takuro Sato}
\IEEEauthorblockA{\hspace{0.6cm}\textit{Faculty of Science and Engineering} \\
\textit{Waseda University}\\
Tokyo, Japan \\
t-sato@waseda.jp}
\and
\IEEEauthorblockN{\hspace{0.5cm}Tetsuya Kawanishi}
\IEEEauthorblockA{\hspace{0.5cm}\textit{Faculty of Science and Engineering} \\
\textit{Waseda University}\\
Tokyo, Japan \\
kawanishi@waseda.jp}
\par
}

\PlaceAcceptedManuscriptNotice
\maketitle

\begin{abstract}
In this study, we conducted Line-of-Sight (LoS) 2$\times$2 multiple-input multiple-output (MIMO) orthogonal frequency-division multiplexing (OFDM) transmission experiments in the 300-GHz band to investigate the potential for channel capacity enhancement. A circularly polarized patch antenna, supporting both cross- and co-polarization, and a 2$\times$2 MIMO transceiver equipped with a Pre-Correction scheme were developed. Experimental evaluation was carried out under conditions of a stream spacing of 0.3 m and a transmission distance of 1.9 m. As a result, in both co- and cross-polarized antenna configurations, the off-diagonal components were suppressed below -20 dB, and 2-stream transmission with 16-QAM modulation achieved a throughput of 13.1 Gbit/s and a spectral efficiency of 6.55 bit/s/Hz. Furthermore, future research directions toward advanced spatial multiplexing based on channel correlation analysis are discussed. These results demonstrate the feasibility of high-capacity fixed wireless links and provide practical design insights for long-distance transmission.
\end{abstract}


\section{Introduction}

In recent years, communication traffic in mobile networks, high-definition video streaming, and industrial automation has been rapidly increasing, and wireless communication systems are also required to provide ultra-high-speed and large-capacity transmission capabilities exceeding 100 Gbit/s \cite{Song2022THzChallenges}. Optical fiber networks have been widely deployed to address such traffic growth; however, their installation is often difficult or costly in certain environments. High-speed and high-capacity fixed wireless links in the sub-terahertz (THz) band are envisioned as a complementary solution to optical fiber.

The sub-THz band, particularly the 300-GHz band, offers a vast frequency spectrum suitable for large-capacity communications. Transmission experiments for limited-area networks such as stadiums, point-to-point access, and mobile backhaul (MBH) have been actively investigated. While the 300-GHz band provides wide bandwidth, its short wavelength and strong directivity result in limited natural scattering and diffraction. Therefore, narrow-beam formation using high-gain antennas is essential. In addition, large free-space path loss and the lack of rich multipath components, unlike the microwave band, present challenges for both indoor and outdoor environments. Although conventional multiplexing techniques that exploit multipath may not be effective, it has been shown that spatial multiplexing via multiple-input and multiple-output (MIMO) is feasible in LoS environments by optimizing antenna placement and beamforming \cite{Eckhardt2023Capacity}. In addition, attempts to enhance capacity using random reflective surfaces (RRSs) have been reported \cite{Singh2024RRS}. However, since reflected paths inevitably involve higher losses, LoS-MIMO is considered an effective design strategy for long-distance communications in the 300-GHz band \cite{Maletic2021LOSSubTHz}.

The authors have previously developed high-gain lens patch antennas and conducted single-stream orthogonal frequency-division multiplexing (OFDM) transmission experiments aimed at extending communication distance \cite{b3,b4}. In this paper, we present a proof-of-concept demonstration of channel capacity enhancement using 300-GHz circularly polarized lens patch antennas supporting beamforming, together with a 2$\times$2 MIMO OFDM transceiver. Specifically, we investigate channel separation based on cross- and co-polarized patch antennas, and evaluate transmission performance in a 2$\times$2 MIMO OFDM environment. These experiments aim not only to verify the feasibility of high-capacity fixed wireless links in the 300-GHz band but also to provide practical design knowledge.

\section{System Overview}
\subsection{Overview of Mobile Network}

\begin{figure}[htbp]
    \centerline{\includegraphics[width=8.2cm, height=4.2cm]{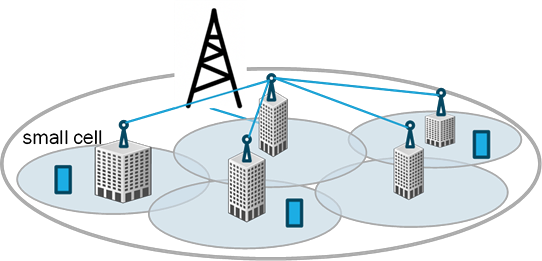}}
    \caption{Overview of Mobile Networks in the RAN.}
    \label{fig:mbh}
\end{figure}

With the introduction of B5G/6G, a significant increase in user data traffic is expected, leading to throughput demands exceeding 100~Gbit/s in Mobile Fronthaul (MFH) and Mobile Backhaul (MBH). Fig.~\ref{fig:mbh} illustrates an overview of mobile networks in the Radio Access Network (RAN).

Particularly, narrow-beam transmission leveraging the high directivity of terahertz waves is promising for suppressing interference and effectively utilizing wideband spectrum resources in higher-frequency bands. Moreover, with the densification of small cells, the number of radio units (RUs) inside Multi-Function Heads (MFHs) is expected to increase. To address this, PHY-Low baseband functions—such as analog/digital beamforming and spatio-temporal processing using fast Fourier transform (FFT)—are being reconsidered, and new implementation approaches in the distribution unit (DU) are under active discussion \cite{b2-1, b2-2, b2-3}.

Thus, sub-THz wireless systems are expected to serve as complementary technologies to optical fiber in future MFH/MBH networks. In particular, medium-range high-capacity transmission using 300-GHz LoS-MIMO will be an important component in B5G/6G network design for MFH/MBH.

\section{300-GHz Band Transceiver Design}
\subsection{Design Strategies for LoS-MIMO Transmission}

In 300-GHz LoS propagation, multipath diversity—commonly seen in the microwave band—cannot be exploited. The spatial independence of channels is therefore determined by antenna placement and geometric path differences. Physical strategies to increase channel rank in LoS-MIMO include suppressing off-diagonal components using cross-polarization, optimizing beam diameter and stream spacing with high-directivity antennas, antenna placement to maximize angle-of-arrival differences, beamforming optimization, and the introduction of artificial reflective surfaces.

The main design considerations for LoS-MIMO transmission in the 300-GHz band targeted in this work are as follows:

\begin{enumerate}
  \item The short wavelength and strong directivity of the 300-GHz band, combined with surface roughness comparable to the wavelength (e.g., concrete walls), result in significant scattering and reflection losses. Hence, communication design assuming LoS propagation is essential.
  \item Due to higher free-space path loss compared with microwave bands, high-gain and narrow-beam antennas such as lens or horn antennas are mandatory.
  \item Based on the use cases discussed in Section II, building-to-building LoS links are expected to be limited to transmission distances below 1 km.
\end{enumerate}

To demonstrate the feasibility of capacity enhancement in LoS-MIMO, we developed circularly polarized patch antennas capable of both co- and cross-polarization, along with a 2×2 MIMO transceiver. Experimental transmission tests and channel correlation analyses were conducted using the developed system.

\subsection{System Design}

System design of LoS-MIMO in the 300-GHz band requires maximizing channel capacity and optimizing stream separation. The relationship between the channel capacity and the channel matrix $\mathbf{H}$ of a general MIMO system is expressed as:

\begin{equation}
\label{eq:capacity}
    C = \log_{2} \det \left( \mathbf{I}_{N_r} + \frac{\rho}{N_t} \mathbf{H}\mathbf{H}^H \right),
\end{equation}

For a 2$\times$2 MIMO system, the channel matrix $\mathbf{H}$ can be represented as:

\begin{equation}
\label{eq:channel-matrix}
    \mathbf{H} =
    \begin{bmatrix}
    h_{11} & h_{12} \\
    h_{21} & h_{22}
    \end{bmatrix},
\end{equation}

where $h_{ij}$ denotes the channel coefficient from transmit antenna $j$ to receive antenna $i$.  
$h_{11}$ and $h_{22}$ are diagonal components, while $h_{12}$ and $h_{21}$ are off-diagonal components.  
The variables are defined as:  
$N_t$: number of transmit antennas,  
$N_r$: number of receive antennas,  
$\rho$: per-antenna average SNR,  
$\mathbf{I}_{N_r}$: $N_r \times N_r$ identity matrix,  
$\mathbf{H}^H$: Hermitian transpose of $\mathbf{H}$,  
$\det(\cdot)$: determinant,  
$C$: MIMO channel capacity [bit/s/Hz].

To increase channel capacity in LoS-MIMO, ensuring high-rank channel matrices is essential, which requires reducing the receive spatial correlation coefficient. This enhances stream independence and thereby channel capacity. The receive spatial correlation coefficient is expressed as:

\begin{equation}
\label{eq:spatial-correlation}
    \rho = \frac{E \!\left[ h_{11} h_{12}^* + h_{21} h_{22}^* \right]}
    {\sqrt{E \!\left[ |h_{11}|^2 + |h_{21}|^2 \right]} \, 
     \sqrt{E \!\left[ |h_{12}|^2 + |h_{22}|^2 \right]}},
\end{equation}

where $\rho$ denotes the spatial correlation coefficient ($|\rho| \to 0$: uncorrelated streams, $|\rho| \to 1$: fully correlated), $E[\cdot]$ indicates time or frequency averaging, and $(\cdot)^*$ denotes complex conjugation. In this paper, we employ $\rho$ as a metric to analyze the independence of MIMO streams.

\subsection{Pre-Correction Scheme}

\begin{figure}[htbp]
    \centerline{\includegraphics[width=8.8cm]{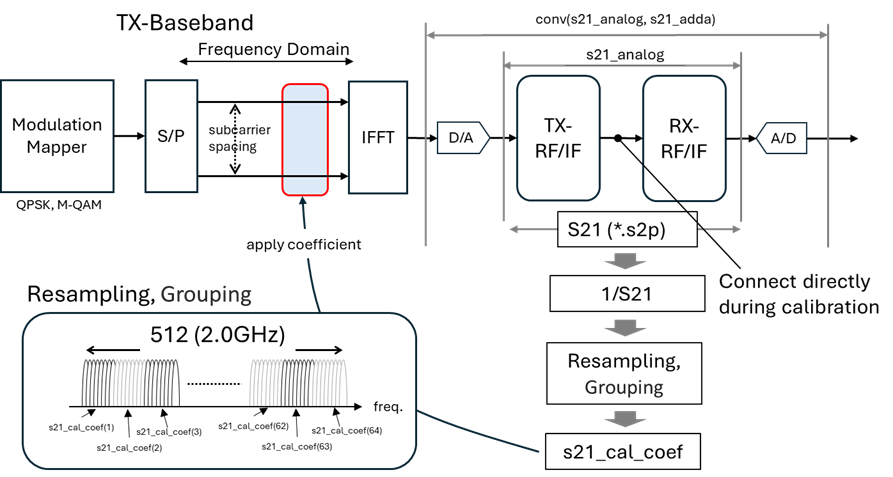}}
    \caption{Block diagram of Pre-Correction.}
    \label{fig:Fig_PreCorrection1}
\end{figure}

Pre-Correction is a technique that compensates for distortions in the analog front-end (including Digital to Analog Converters (DAC) and Analog to Digital Converter (ADC)) at the transmitter side. Fig.~\ref{fig:Fig_PreCorrection1} shows the block diagram of the scheme. During calibration, the RF and IF sections of the TX and RX chains of each stream were connected directly, and the S21 characteristics were measured with a vector network analyzer (VNA). Baseband S21 characteristics including DAC and ADC were measured separately and combined via convolution to obtain the overall response. An inverse response was then calculated, resampled to 64 points matching the subcarrier spacing, and multiplied with each OFDM subcarrier in the frequency domain prior to IFFT. This procedure effectively flattens the long preamble and OFDM payload in the frequency response after ADC conversion.

Alternatively, a direct channel estimate obtained from a loopback connection can also be loaded into a transmitter-side equalizer to achieve a similar compensation effect. In this work, we adopted the Pre-Correction scheme to reduce circuit complexity.

\begin{figure}[htbp]
    \centerline{\includegraphics[width=8.8cm]{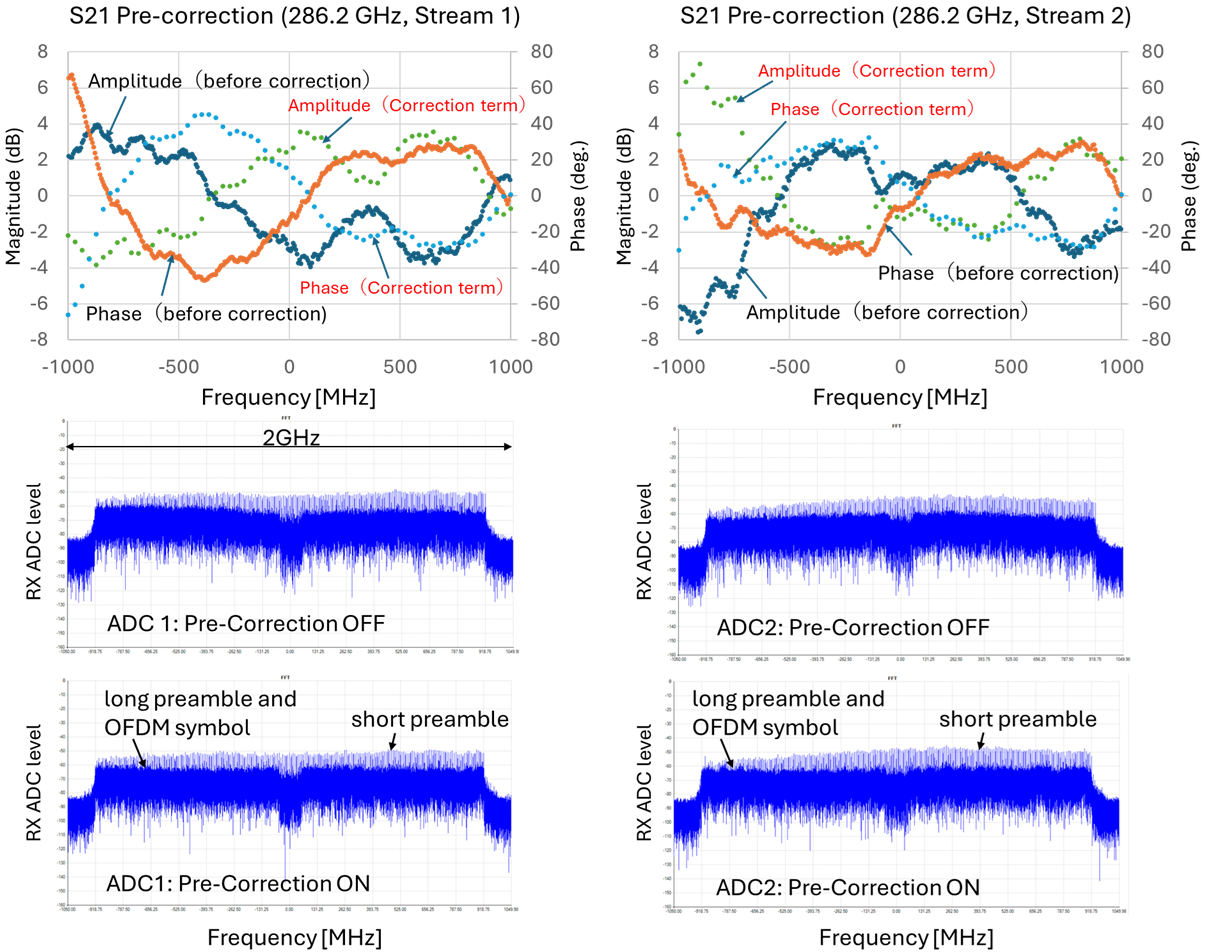}}
    \caption{Effectiveness of Pre-Correction (Top: Frequency Response of Analog Front-End, Bottom: Improved ADC Characteristics After Pre-Correction).}
    \label{fig:Fig_PreCorrection}
\end{figure}

Fig.~\ref{fig:Fig_PreCorrection} shows an example result of applying Pre-Correction. Measurements were taken with 501 points using a VNA, downsampled to 64 for frequency-domain equalization. By compensating fixed analog distortion at the transmitter, noise enhancement in receiver equalization is mitigated. As a result, the receive SNR improved. The right-hand chart shows that the OFDM spectrum across a 2-GHz bandwidth was equalized within ±1 dB, confirming SNR improvement.

\subsection{Baseband Design}

The baseband configuration is identical to that used in our previous SISO-OFDM transmission experiments \cite{b4}. The occupied bandwidth is 2~GHz, and a 512-subcarrier OFDM scheme (512-OFDM) is adopted. The OFDM symbol length is 290~ns, with a cyclic prefix (CP) of 34~ns. The time-division duplex (TDD) frame consists of a short preamble, two long preambles, followed by 20 OFDM symbols (data payload). The preamble employs Zadoff-Chu sequences.

\begin{figure}[htbp]
    \centerline{\includegraphics[width=8.8cm]{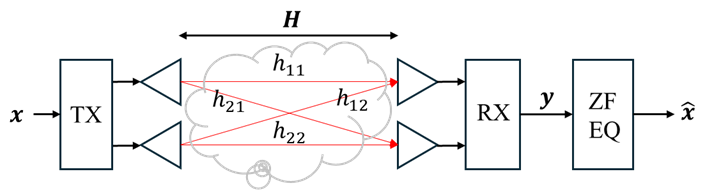}}
    \caption{Conceptual transmission model of 2$\times$2 MIMO.}
    \label{fig:Fig_zf-eq}
\end{figure}

To describe the receiver equalization, Fig.~\ref{fig:Fig_zf-eq} shows the 2$\times$2 MIMO transmission model including the baseband section. The estimation of the transmitted vector using a zero-forcing (ZF) equalizer is given in (\ref{eq:zf-eq1})-(\ref{eq:zf-eq2}):

\begin{equation}
    \label{eq:zf-eq1}
    \mathbf{y} = \mathbf{H} \mathbf{x} + \mathbf{n}
\end{equation}

\begin{equation}
    \label{eq:zf-eq2}
    \widehat{\mathbf{x}} = (\mathbf{H}^H \mathbf{H})^{-1} \mathbf{H}^H \mathbf{y},
\end{equation}

where $\mathbf{H}$ is the MIMO channel matrix defined in (\ref{eq:channel-matrix}),  
$\mathbf{y} \in \mathbb{C}^{N_r \times 1}$ is the received signal vector,  
$\mathbf{x} \in \mathbb{C}^{N_t \times 1}$ is the transmitted signal vector, and  
$\mathbf{n} \in \mathbb{C}^{N_r \times 1}$ is the noise vector.  

In ZF equalization, the ZF weight $(\mathbf{H}^H \mathbf{H})^{-1} \mathbf{H}^H$ is applied to estimate $\widehat{\mathbf{x}}$ from the received vector $\mathbf{y}$.

Channel estimation is performed by transmitting long preambles from TX1 and TX2 in a time-division manner, allowing simultaneous reception at RX1 and RX2 for full 2$\times$2 MIMO estimation.

Equalization is carried out sequentially using two steps:  
(1) channel estimation and equalization with the long preamble,  
(2) channel estimation and equalization with pilot subcarriers.  

In the second step, pilot subcarriers inserted every 16 subcarriers provide amplitude/phase variations, which are interpolated using an oversampled Hilbert filter and applied to equalize all subcarriers.

\section{Experimental Results and Discussion}
\subsection{Experimental Setup}

We constructed an experimental setup for 2$\times$2 MIMO-OFDM transmission at 286.2~GHz to evaluate channel characteristics with a stream spacing of 0.3~m, transmission distance of 1.9~m, and antenna height of 1.5~m. The main experimental parameters are listed in Table~\ref{table:parameters}.

In our prior SISO-OFDM experiments \cite{b4}, transmission distance of 72.4~m was achieved using transmit power of 4.04~dBm and total antenna gain of 79.6~dBi. By contrast, in the present MIMO experiment, the power amplifier configuration was modified. With an RF PA of Psat 10.5~dBm, the effective transmit power was 2~dBm. The total antenna gain was 52~dBi, and the link distance was reduced to 1.9~m. Thus, compared with the SISO case, both the transmit power and antenna gain sum were about 30~dB lower, and the transmission distance was about $1/32$. This configuration was chosen intentionally to avoid overly small off-diagonal terms and to evaluate stream separation.

At the transmitter, 16~dBic RHCP and LHCP patch antennas with HPBW 18° were used, while the receiver employed lensed RHCP/LHCP patch antennas with 36~dBic gain and HPBW 1.7°. Both co-polarized (RHCP-RHCP) and cross-polarized (RHCP-LHCP) stream pairs were evaluated. The transmit RF power was about 2~dBm (after beamforming), with a carrier frequency of 286.2~GHz and 512-OFDM modulation (bandwidth 2~GHz). QPSK and 16-QAM were employed as modulation schemes.

\begin{figure}[htbp]
    \centerline{\includegraphics[width=8.8cm]{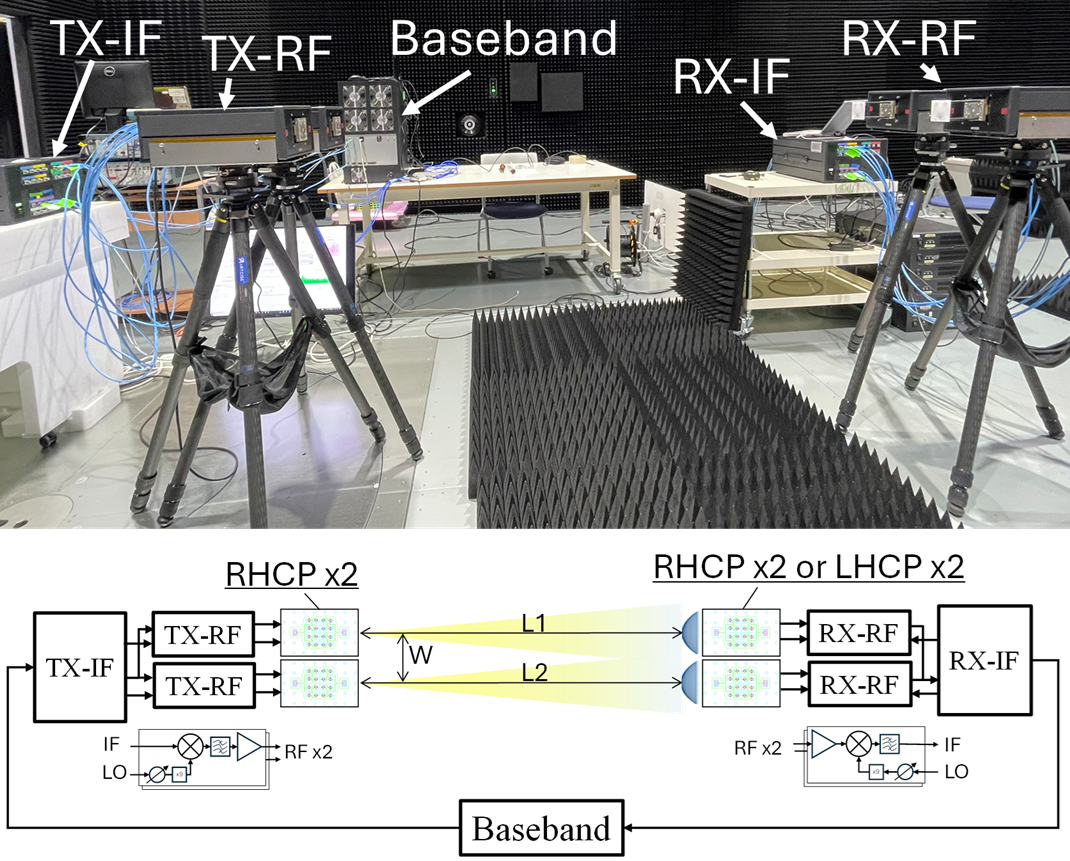}}
    \caption{Experimental setup (top) and transceiver block diagram (bottom).}
    \label{fig:experimental_view}
\end{figure}

Pre-Correction was applied at the transmitter to compensate for analog front-end distortions, ensuring balanced amplitude and phase across streams. Experiments were conducted in an anechoic chamber. Fig.~\ref{fig:experimental_view} shows the experimental view and transceiver block diagram. The EIRP was 18~dBm, free-space path loss over 1.9~m was 87~dB, and thermal noise at 2~GHz bandwidth and 297~K was -80.8~dBm. With receiver noise figure of 15~dB, the received power was estimated at -69~dBm. Since the per-subcarrier OFDM signal is about 27~dB below CW, the per-subcarrier SNR was estimated at 11.8~dB.

\begin{table}[htbp]
    \caption{Experimental Parameters.}
    \label{table:parameters}
    \centering
    \begin{tabular}{p{1.2cm}||p{6.8cm}}
        \hline
        Tx-Rx & Stream Spacing (W): 0.300 m \\
        Allocation & Height of Tx and Rx (H): 1.500 m \\
        & Tx-Rx Distance (L1)(L2): 1.90 m \\
        \hline
        & Transmitter: \\
        & \hspace{0.1cm}2-port RHCP Patch Antenna, 16 dBic (HPBW 18°) \\
        Antennas & \hspace{0.1cm}2-port LHCP Patch Antenna, 16 dBic (HPBW 18°) \\
        & Receiver: \\
        & \hspace{0.1cm}2-port Lensed RHCP Patch Antenna, 36 dBic (HPBW 1.7°) \\
        & \hspace{0.1cm}2-port Lensed LHCP Patch Antenna, 36 dBic (HPBW 1.7°) \\
        \hline
        & Carrier Frequency: 286.2 GHz \\
        & Occupied Bandwidth: 2.0 GHz \\
        & RF Power: 2 dBm (approx., with beamforming) \\
        Transceiver & Transmission Method: 2$\times$2 MIMO-OFDM (512-OFDM) \\
        & OFDM Symbol Length: 290 ns (incl. CP 34 ns) \\
        & Modulation: QPSK, 16-QAM \\
        & Equalizer: ZF (Preamble, Pilot Subcarrier) \\
        \hline
    \end{tabular}
\end{table}

\subsection{Results and Discussion}

In the 286.2~GHz 2$\times$2 MIMO-OFDM setup, channel characteristics and receive SNR were evaluated. Under the test condition (1.9~m, HPBW=1.7°), the receive beam diameter was 0.056~m, while the stream spacing was 0.3~m, yielding about 5.3$\times$ separation.

\begin{figure}[htbp]
    \centerline{\includegraphics[width=8.0cm]{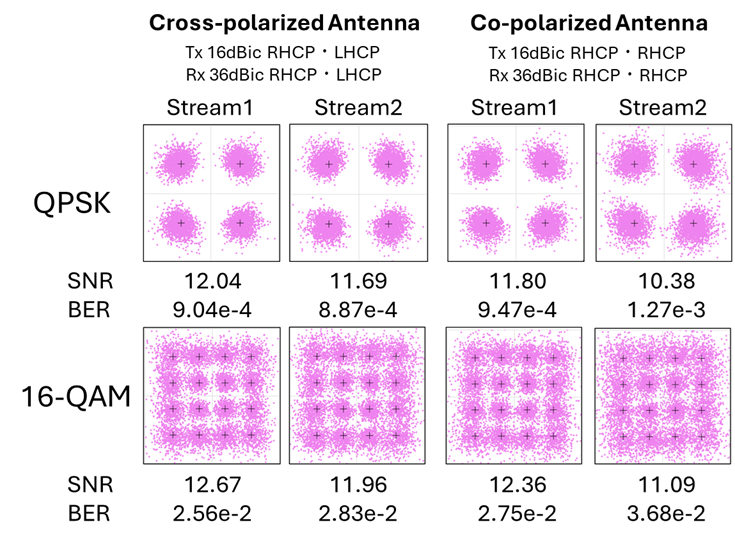}}
    \caption{Receiver performance for 2$\times$2 MIMO-OFDM transmission.}
    \label{Fig:const_1}
\end{figure}

Fig.~\ref{Fig:const_1} shows receiver performance. Both co-polarized and cross-polarized setups yielded similar SNR, confirming stable operation due to narrow-beam formation by the lensed antennas. The measured SNR was about 12~dB, consistent with the design, despite intentional suppression of antenna gain to evaluate off-diagonal components. With 16-QAM, throughput of 13.1~Gbit/s and spectral efficiency of 6.55~bit/s/Hz were achieved. Under the same conditions, the theoretical spectral efficiency from (\ref{eq:capacity}) with SNR=11.8~dB and rank-2 $\mathbf{H}$ is 6.20~bit/s/Hz, matching well with the experimental result. These results suggest the potential for further stream multiplexing by combining co- and cross-polarization.

\begin{figure}[htbp]
    \centerline{\includegraphics[width=8.8cm]{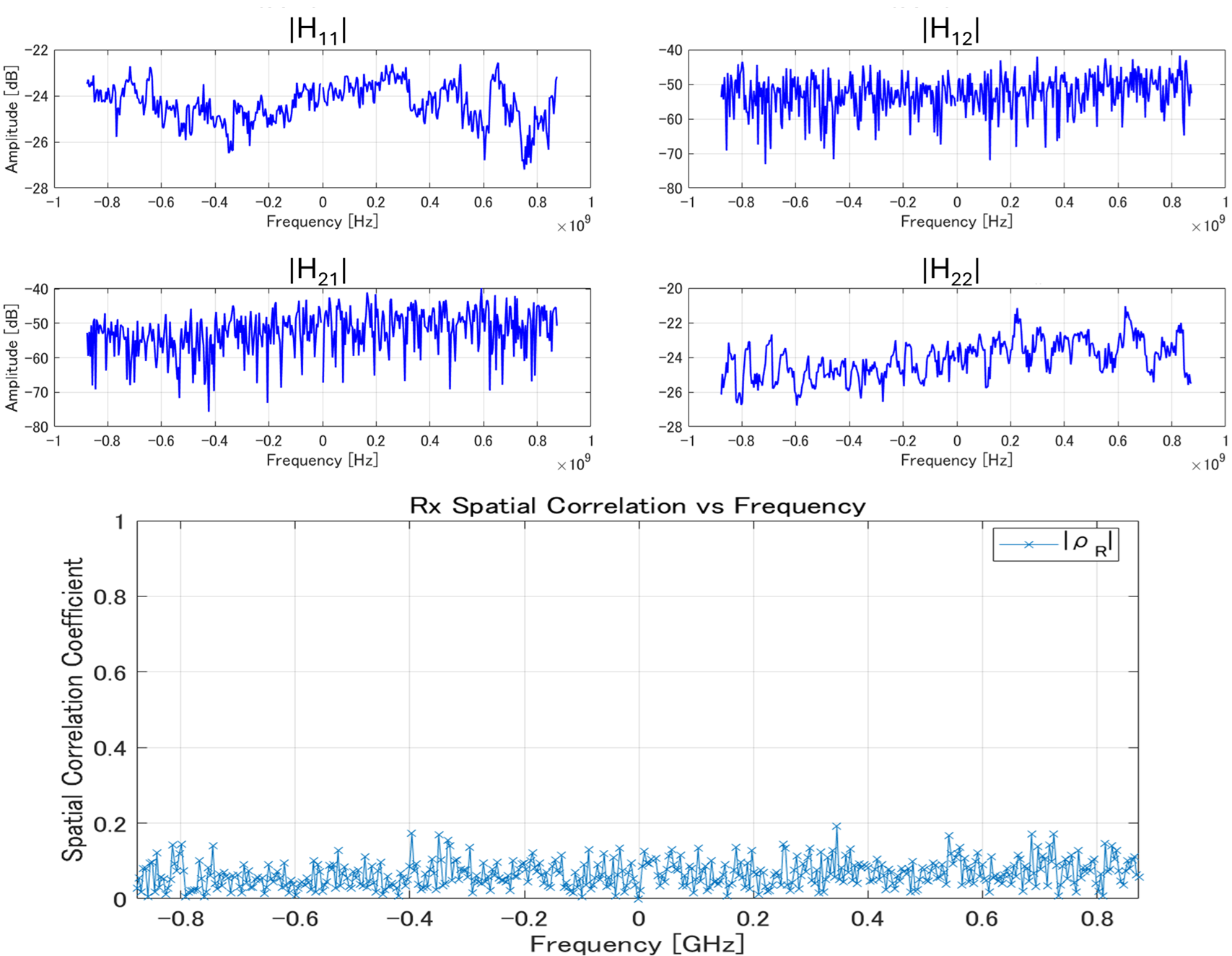}}
    \caption{Channel characteristics for cross-polarized antennas. Top: frequency response of channel $H$; bottom: Rx spatial correlation.}
    \label{fig:Fig_Chest_cross-pol}
\end{figure}

\begin{figure}[htbp]
    \centerline{\includegraphics[width=8.8cm]{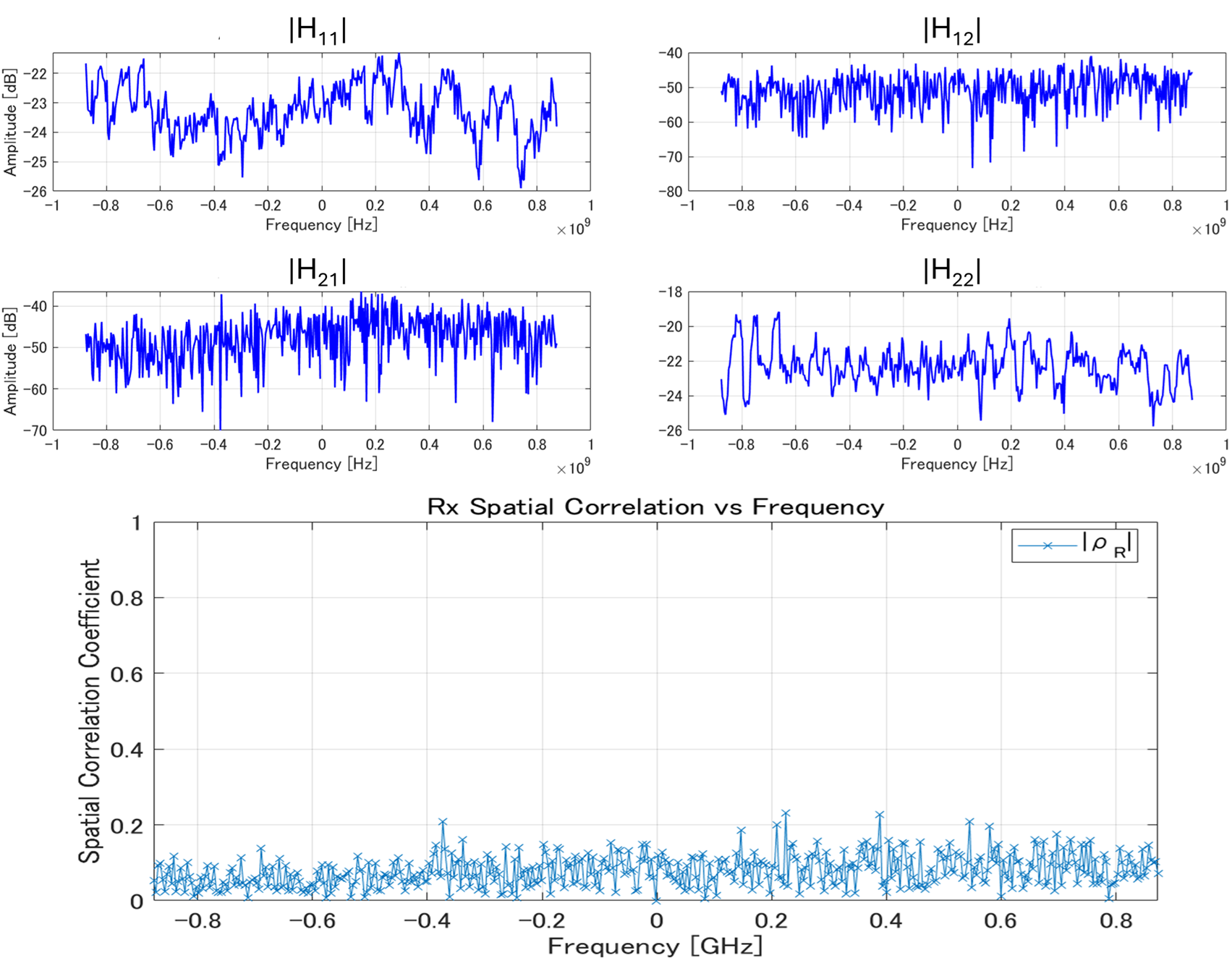}}
    \caption{Channel characteristics for co-polarized antennas. Top: frequency response of channel $H$; bottom: Rx spatial correlation.}
    \label{fig:Fig_Chest_co-pol}
\end{figure}

Figs.~\ref{fig:Fig_Chest_cross-pol} and \ref{fig:Fig_Chest_co-pol} show channel responses for cross- and co-polarization. The upper plots show frequency response of channel matrix $\mathbf{H}$, while the lower plots show receive spatial correlation. Off-diagonal components $h_{12}$ and $h_{21}$ were suppressed below -20~dB for both cases, indicating effective stream separation. The spatial correlation coefficient was below 0.1, confirming sufficiently low correlation.

We further examined scaling to long-distance transmission (1~km). Compared with prior SISO-OFDM experiments \cite{b4}, extending from 72.4~m to 1~km requires an additional 22.7~dB antenna gain sum, for a total of at least 102~dBi. Assuming 55~dBi antennas at both ends and antenna efficiency $\eta$ = 0.5, HPBW is about 0.26°, and the beam diameter at 1~km is about 4.5~m. To maintain separation comparable to the short-range experiment, stream spacing of approximately 24~m is required.

Thus, simply opposing high-gain antennas is insufficient for long-distance stream separation. Techniques such as lens arrays, phase control, and slanted offset placement of transmit/receive streams to introduce angular separation, as well as reflective surfaces (RRS), are promising strategies.

\section{Conclusion}

We developed and tested a 286.2~GHz 2$\times$2 MIMO-OFDM transceiver system, evaluating both channel and transmission performance. The results demonstrated 16-QAM dual-stream transmission achieving 13.1~Gbit/s throughput and 6.55~bit/s/Hz spectral efficiency. Narrow-beam formation by the lensed antennas enabled stable performance with both co- and cross-polarization.

Furthermore, based on the results, we investigated scaling to long-distance transmission. Maintaining equivalent stream separation at 1~km requires a spacing of about 24~m. Therefore, simple high-gain antenna opposition is insufficient. Future work should focus on phase control, slanted offset stream placement, and RRS introduction to improve orthogonality between beams and enable scalable LoS-MIMO links.

\section*{Acknowledgment}
These research results were obtained from the commissioned research (JPJ012368C00302, JPJ012368C00491) by the National Institute of Information and Communications Technology (NICT), and the ASPIRE program (JPMJAP2324) of the Japan Science and Technology Agency (JST), Japan.


\vspace{12pt}

\end{document}